\documentstyle[amsfonts,aps,twocolumn,epsfig]{revtex}
\begin{document}
\title{\begin{flushright}
\footnotesize{CECS-PHY-00/11}\\
\end{flushright}{\vskip 1.0cm}Black holes with topologically nontrivial AdS
asymptotics}
\author{Rodrigo Aros$^{1,2}$, Ricardo Troncoso$^{1}$ and Jorge Zanelli$^{1,3}$}
\address{$^{1}$Centro de Estudios Cient\'{\i }ficos (CECS), Casilla 1469,
Valdivia, Chile.\\
$^{2}$Universidad Nacional Andres Bello, Sazie 2320, Santiago,Chile \\
$^{3}$Departamento de F\'{\i }sica, Universidad de Santiago de Chile,Casilla 307, Santiago 2,
Chile.}
\date{November 13, 2000}
\maketitle
\begin{abstract}
Asymptotically locally AdS black hole geometries of dimension $d\geq 3$ are studied for nontrivial
topologies of the transverse section. These geometries are static solutions of a set of theories
labeled by an integer $k\in \{1,2,...,[(d-1)/2]\}$ which possess a unique globally AdS vacuum. The
transverse sections of these solutions are $d-2$ surfaces of constant curvature $\gamma $,
normalized to $\gamma =\pm 1,0$ allowing for different topological configurations. The
thermodynamic analysis of these solutions reveals that the presence of a negative cosmological
constant is essential to ensure the existence of stable equilibrium states. In addition, it is
shown that these theories are holographically related to $[(d-1)/2]$ different conformal field
theories at the boundary.

PACS number(s): 04.50.+h, 04.20.Jb, 04.70.Dy
\end{abstract}

\section{Introduction}

Several formal arguments suggest that at a fundamental level, the cosmological constant
($\Lambda$) should be negative, in spite of recent observations that favor a positive effective
value at a cosmic scale. As shown by Hawking and Page \cite{Hawking:1983dh} the presence of a
negative $\Lambda$ makes it possible for black holes to reach thermal equilibrium with a heat
bath. Moreover, a negative cosmological constant is required to obtain a correct definition of
Noether charges representing the mass and angular momentum \cite{Aros:1999id,Aros:1999kt}. Even in
a spacetime with vanishing $\Lambda$, the correct results are obtained provided the construction
is carried out with $\Lambda<0$ and taking the $\Lambda \rightarrow 0$ limit at the end. In this
sense, $\Lambda$ acts as a regulator allowing the canonical ensemble as well as Noether charges,
to be well defined.

On the other hand, black holes with $\Lambda >0$ do not admit a global definition of time, which
prevents the existence of a positive energy theorem. This fact is related with the nonexistence of
a locally supersymmetric extension of gravity for positive $\Lambda$ \cite{Abbott:1982ff}.

The {\it topological censorship theorem } \cite{Friedman:1993ty} states that in asymptotically flat
spacetimes only spherical horizons can give rise to well defined causal structure for a black hole.
This is circumvented by the presence of a negative cosmological constant, in which case well
defined black holes with locally flat or hyperbolic horizons have been shown to exist
\cite{Lemos:1995xp,Vanzo:1997gw,Brill:1997mf}. This kind of black holes with topologically
nontrivial AdS asymptotics are relevant in testing the AdS/CFT correspondence \cite{Aharony:1999ti}
in the cases where the thermal CFT is defined on backgrounds of different topologies such as $
S^{1}\times {\Bbb R}^{d-2}$, $S^{1}\times S^{d-2}$ or $S^{1}\times H^{d-2}$ \cite{Witten:1998zw},
\cite{Birmingham:1998nr,Emparan:1999pm}. As will be shown, these backgrounds correspond to the
asymptotic regions of the solutions discussed below.

As observed in \cite{Balasubramanian:2000wv} the renormalization group equation for a CFT at the
boundary should be obtained from the radial Hamiltonian constraint of a gravitation theory with
higher powers of the curvature.

In dimensions higher than four, in addition to Einstein's theory, there exist a host of sensible
gravity actions which contain higher powers of the curvature (for a recent review see e.g.
\cite{HDG} and references therein). A particular class of these theories gives rise to second
order field equations for the metric, which possess spherical black hole solutions with a well
defined asymptotically AdS behavior \cite{Crisostomo:2000bb}. In this work, that family of black
hole solutions is extended to geometries with locally flat and hyperbolic horizons, including
different non trivial asymptotic topologies. The thermal behavior of this class of theories
around these solutions is analyzed in detail and are shown to be holographically connected with
different thermal CFT's.

\section{AdS gravity theories in higher dimensions}

A minimally sensible gravity theory should be described by an action leading to second order field
equations for the metric in order to avoid problems with causality in the classical theory, or
ghosts at the quantum level. The so called Lanczos-Lovelock gravities are the only purely metric
theories that fulfill this requirement \cite{LL}, although they present several problems. For each
dimension, they possess a number of arbitrary constants. As a consequence, their field equations
allow the existence of spherically symmetric black holes with negative energy, as well as positive
energy solutions with naked singularities. Moreover, these solutions do not have a unique
asymptotic behavior and can even spontaneously jump between distinct geometries
\cite{Wheeler:1986nh,Boulware:1985wk,Teitelboim:1987}.

These problems can be overcome by requiring the theories to possess {\em a unique} cosmological
constant, which strongly restricts the arbitrary coefficients in the Lanczos-Lovelock actions.
Hence one is led to a set of theories which, for each dimension $d$, have a number of Lagrangians
labeled by an integer $k$ which represents the highest power of curvature in the Lagrangian. The
action reads
\begin{equation}
I_{k}=\kappa \int \sum_{p=0}^{k}c_{p}^{k}L^{(p)}\;,  \label{Ik}
\end{equation}
where $L^{(p)}$ is given by
\begin{equation}
L^{(p)}=\epsilon _{a_{1}\cdots a_{d}}R^{a_{1}a_{2}}\!\cdot \!\cdot \!\cdot
\!R^{a_{2p-1}a_{2p}}e^{a_{2p+1}}\!\cdot \!\cdot \!\cdot \!e^{a_{d}}, \label{Lovlag}
\end{equation}
and
\begin{equation}
c_{p}^{k}=\left\{
\begin{array}{ll}
\frac{l^{2(p-k)}}{(d-2p)}\left(
\begin{array}{c}
k \\
p
\end{array}
\right)  & \text{ }p\leq k \\
0 & \text{ }p>k
\end{array}
\right.,   \label{Coefs}
\end{equation}
with
\begin{equation}\label{k}
1\leq k\leq [\frac{d-1}{2}],
\end{equation}
where $[x]$ stands for the integer part of $x$.

Unlike a generic Lanczos-Lovelock theory, which would be obtained for arbitrary coefficients
$c_{p}^{k}$, the action $I_{k}$ possesses only two fundamental constants, $\kappa $ and $l$,
related to the gravitational constant $G_{k}$ and the cosmological constant $\Lambda $
through\footnote{ The gravitational constant has natural units given by $[G_{k}]=($length$
)^{d-2k}$, and $l$ corresponds to the AdS radius. In this work  the conventions of Ref.
\cite{Crisostomo:2000bb} are followed.}
\begin{eqnarray}
\kappa &=&\frac{1}{2(d-2)!\Omega _{d-2}G_{k}},  \label{Kappa} \\
\Lambda &=&-\frac{(d-1)(d-2)}{2l^{2}},  \label{Lambda}
\end{eqnarray}
where $\Omega _{d-2}$ is the surface area of a unit $d-2$-sphere.

The field equations read
\begin{eqnarray}
\epsilon _{ba_{1}\cdots a_{d-1}}\bar{R}^{a_{1}a_{2}}\!\cdot \!\cdot \!\cdot
\!\bar{R}^{a_{2k-1}a_{2k}}e^{a_{2k+1}}\!\cdot \!\cdot \!\cdot \!e^{a_{d-1}}
&=&0,  \label{kEinstein} \\
\epsilon _{aba_{3}\cdots a_{d}}\bar{R}^{a_{3}a_{4}}\!\cdot \!\cdot \!\cdot \!
\bar{R}^{a_{2k-1}a_{2k}}T^{a_{2k+1}}e^{a_{2k+2}}\!\cdot \!\cdot \!\cdot \!e^{a_{d-1}} &=&0,
\label{kTorsion}
\end{eqnarray}
where $\bar{R}^{ab}:=R^{ab}+\frac{1}{l^{2}}e^{a}e^{b}$ and $T^{a}$ is the torsion $2$-form.

Note that the Einstein-Hilbert action in $d$ dimensions is obtained by setting $k=1$ in
Eq.(\ref{Ik}). This is the only possible choice in $3$ and $4$ dimensions, while in five or more
dimensions there are other inequivalent theories with $k \geq 2$. In the case $d=5$ and $k=2$ the
Lagrangian can be cast as the Euler-Chern-Simons form for the AdS group \cite{Chamseddine:1989nu}.
The Euler-Chern-Simons form is obtained from (\ref{Ik}) in any odd dimension for the maximum
allowed value $k=\frac{d-1}{2}$. The locally supersymmetric extension of these last kind theories
with negative cosmological constant is known to exist for any odd dimension, in particular for
$d=11$ \cite{Troncoso:1998va}.

An illustrative example of the kind of theories considered here, is the $k=2$ case. This action,
which exists only for $d>4$, is written in tensor components as
\begin{equation}
I_{2}=\frac{-2(d-3)!\kappa }{l^{2}} \int d^{d}x\sqrt{-g} \left[ \frac{l^{2}{\frak
R}^{2}}{2(d-3)(d-4)}+R-\Lambda \right] , \label{GB+}
\end{equation}
where ${\frak R}^{2}$ stands for the Gauss-Bonnet density,
\begin{equation}
{\frak R}^{2}:=(R_{\mu \nu \alpha \beta }R^{\mu \nu \alpha \beta }-4R_{\mu \nu }R^{\mu \nu
}+R^{2}).  \label{EGB}
\end{equation}

In this paper, only torsion-free solutions will be considered, so that Eq. (\ref {kTorsion}) is
trivially satisfied.

\section{Topological Solutions}

\label{TopologicalSolutions}

Let us consider $d$-dimensional static spacetimes whose spatial sections are foliated along the
radial direction by $(d-2)$-dimensional transverse surfaces $\Sigma _{\gamma }$ of constant
curvature $\gamma$. In terms of Schwarzschild-like coordinates, the metric can be written as
\begin{equation}
ds^{2}=-N^{2}(r)f^{2}(r)dt^{2}+\frac{dr^{2}}{f^{2}(r)}+r^{2}d\sigma _{\gamma
}^{2},  \label{ds}
\end{equation}
where $-\infty <t<\infty $, and $0\leq r<\infty $ is the radial coordinate for which
$r\rightarrow \infty $ defines the asymptotic region. The arc length $d\sigma _{\gamma }^{2}$
corresponds to the distance on $\Sigma _{\gamma }$. Substituting the ansatz (\ref{ds}) in the
field equations (\ref {kEinstein}), leads to the following equations for $N(r)$ and $f^{2}(r)$,
\begin{eqnarray}
\frac{dN}{dr} &=&0,  \label{EqnGravN} \\
\frac{d}{dr}\left( r^{d-1}\left[ F_{\gamma }(r)+\frac{1}{l^{2}}\right]
^{k}\right) &=&0,  \label{EqnGravF}
\end{eqnarray}
where the function $F_{\gamma }(r)$ is given by
\begin{equation}
F_{\gamma }(r)=\frac{\gamma -f^{2}(r)}{r^{2}}.  \label{F(r)}
\end{equation}
By virtue of Eqs. (\ref{EqnGravN}) and (\ref{EqnGravF}), $N$ is a constant,
which can be chosen as $1$ (see Appendix \ref{HamiltonianCharges}), and
\begin{equation}
f^{2}(r)=\gamma +\frac{r^{2}}{l^{2}}-\alpha \left( \frac{2\mu G_{k}}{
r^{d-2k-1}}\right) ^{\frac{1}{k}}  \label{Delta}
\end{equation}
respectively, where $\alpha =(\pm 1)^{k+1}$.

The constant $\gamma $ can be normalized to ${\pm 1,0}$ by an appropriate rescaling of the
coordinates. Thus, the local geometry of $\Sigma _{\gamma }$ is a sphere, a plane\footnote{ For
$\gamma =0$ it is necessary that at least one direction of $\Sigma _{0}$ be compact, otherwise the
integration constant $\mu $ could be rescaled away. } or a hyperboloid,
\[
\Sigma _{{\gamma }}\text{ locally }{=}\left\{
\begin{array}{ll}
S^{d-2} & {\gamma =}\text{ }1 \\
{\Bbb R}^{d-2} & {\gamma =}\text{ }0 \\
H^{d-2} & {\gamma =}-1
\end{array}
\right. .
\]

The solution (\ref{Delta}) may vanish for some values of $r$. The largest
zero of $f^{2}(r)$ corresponds to the outer horizon $r_{+}$ which allows to
express the integration constant $\mu $ as
\begin{equation}
\mu =\frac{r_{+}^{(d-2k-1)}\left( \gamma +\frac{r_{+}^{2}}{l^{2}}\right)
^{k} }{2G_{k}},  \label{mu}
\end{equation}
and is related to the mass $M$ through
\begin{equation}
\mu =\frac{\Omega _{d-2}}{\Sigma _{d-2}}M+\frac{1}{2G_{k}}\delta
_{d-2k,\gamma }.  \label{mass}
\end{equation}
Here $\Sigma _{d-2}$ is the volume of the transverse space, and $\Omega _{d-2}$ corresponds to the
volume of $S^{d-2}$. Note that the mass is shifted with respect to the integration constant $\mu
$\ only for $d-2k=1=\gamma $, which corresponds to a spherically symmetric solution in
Chern-Simons ({\bf CS}) theory (See Appendix \ref{HamiltonianCharges}). Summarizing, for a fixed
value of the label $k$\ in $d$-dimensions, the action $I_{k}$ in Eq. (\ref{Ik}) is extremized by
the metric
\begin{eqnarray}
ds^{2} &=&-\left( \gamma +\frac{r^{2}}{l^{2}}-\alpha \left( \frac{2G_{k}\mu
}{r^{d-2k-1}}\right) ^{1/k}\right) dt^{2}+  \nonumber \\
&&\frac{dr^{2}}{\left( \gamma +\frac{r^{2}}{l^{2}}-\alpha \left( \frac{
2G_{k}\mu }{r^{d-2k-1}}\right) ^{1/k}\right) }+r^{2}d\sigma _{\gamma }^{2}\;,
\label{BHGeneral}
\end{eqnarray}
with $\alpha =(\pm 1)^{k+1}$, whose asymptotic behavior is locally AdS, for any topology of
$\Sigma _{\gamma }$.

Note that the $\mu =0$ solution
\begin{equation}
ds^{2}=-\left( \gamma +\frac{r^{2}}{l^{2}}\right) dt^{2}+\frac{dr^{2}}{ \left( \gamma
+\frac{r^{2}}{l^{2}}\right) }+r^{2}d\sigma _{\gamma }^{2}\;, \label{massless}
\end{equation}
is a locally AdS manifold, which is a common solution to the Eqs. (\ref{kEinstein}) and
(\ref{kTorsion}) for any value of $k$, in particular for Einstein's theory ($k=1$). As discussed
in \cite{Vanzo:1997gw,Brill:1997mf}, when the transverse section is locally hyperbolic ($\gamma
=-1$), although the metric (\ref{massless}) possesses a horizon at $r_{+}=l$ , it may not
describe a black hole. If the transverse section $\Sigma _{-1}$ has topology ${\Bbb R}^{d-2}$,
Eq. (\ref{massless}) is not a black hole, but it could be one provided suitable identifications
are performed on $\Sigma _{-1}$, analogous to the BTZ solution \cite{Banados:1993gq},
\cite{ThanksVanzoandLouko}.

Note that for theories with odd $k$, the line element (\ref{BHGeneral}) is real for all values of
the integration constants, however for even $k$ only positive $\mu $ is allowed. In what follows
it is shown that this metric describes black holes if $f^{2}(r)$ has at least one zero, and they
are naked singularities otherwise. It is apparent from (\ref{BHGeneral}) that the theory with
$d-2k-1=0$ must be treated separately.

\subsection{Generic theories: $d-2k\neq 1$}

In Fig.1, the zeros of $f^{2}(r)$ correspond to the intersections of the parabolas $\left( \gamma
+\frac{r^{2}}{l^{2}}\right) $ and the functions $\alpha \left( \frac{2G_{k}\mu
}{r^{d-2k-1}}\right) ^{1/k}$, for $\gamma =0,\pm 1$, and different values of $\alpha $ and $\mu $
respectively.

\begin{figure}[tbm]
\begin{center}
  \leavevmode
  \epsfxsize=3 in
\epsfbox{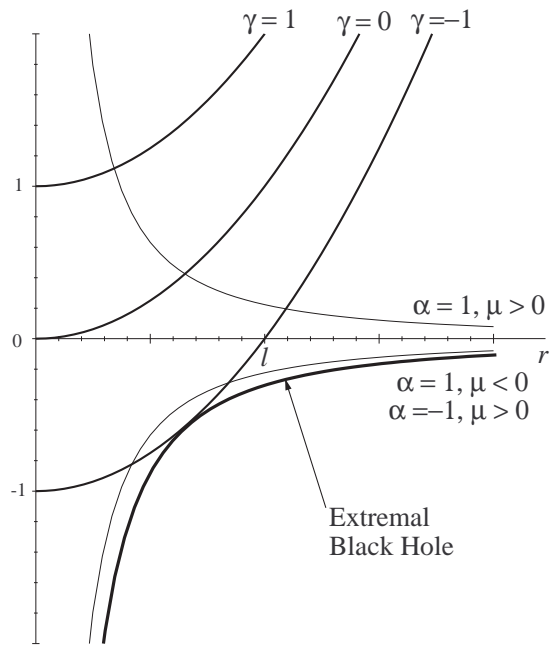}\label{horizons}\caption{The horizons are located at the zeros of
$f^{2}(r)$, which occur at the intersections of the parabolas $\left( \gamma
+\frac{r^{2}}{l^{2}}\right)$ and the functions $\alpha \left(
\frac{2G_{k}\mu}{r^{d-2k-1}}\right)^{1/k}$. These are displayed for $\gamma=0,\pm 1$, and
different values of $\alpha$ and $\mu$. There exist a single horizon for $\alpha=1$ and $\mu \geq
0$. Two horizons arise either for $\alpha=1$ and $\mu_{c}<\mu<0$, or for $\alpha=-1$ and
$0<\mu<\mu_{c}$. In the extreme case, both horizons coalesce for $\mu=\mu_{c}$.}
\end{center}
\end{figure}

It is necessary to consider separately the theories of even and odd $k$. The range of $\mu $ for
which black holes exist is summarized for different values of $\gamma $ and $k$ in the following
table. \vspace{0.5cm}
\begin{center}
{\large Generic theories $(d-2k\neq 1)$}
\end{center}
\begin{equation}
\begin{array}{|l|c|c|c|}
\hline & \gamma =1 & \gamma =0 & \gamma =-1 \\ \hline &  &  &  \\ \text{odd }k\text{ }(\alpha =1)
& \mu > 0 & \mu > 0 &
\begin{array}{ll}
\mu \geq \mu _{c} & :\mu _{c}<0
\end{array}
\\
&  &  &  \\ \hline \text{even }k\text{ } & \mu > 0 & \mu > 0 &
\begin{array}{ll}
\alpha =1 & :\mu \geq 0 \\ &  \\ \alpha =-1 & :\mu _{c}\geq \mu \geq 0
\end{array}
\\ \hline
\end{array}
\label{MassTable}
\end{equation}

For $\gamma =-1$ there is a critical value $\mu _{c}$ given by
\begin{equation}
\mu _{c}=\frac{(-1)^{k}l^{d-2k-1}}{2G_{k}}\sqrt{\frac{
(d-2k-1)^{d-2k-1}(2k)^{2k}}{(d-1)^{d-1}}},  \label{criticalmu}
\end{equation}
which separates topological black holes with hyperbolic transverse section
from naked singularities.

${\bf \bullet \;\gamma =1}$:

This case corresponds to the static, spherically symmetric black holes analyzed in Ref.
\cite{Crisostomo:2000bb}. They have a single event horizon, provided $\mu >0$. For any integer
$k$ such that $1\leq k<(d-1)/2$, these solutions share a common causal structure with the
Schwarzschild-AdS$_{4}$ black hole solution of Einstein's theory ($k=1$).

${\bf \bullet \;\gamma =0}$:

In this case, the set of geometries described by (\ref{BHGeneral}) possess locally flat
transverse sections $\Sigma _{0}$, which are assumed to be orientable. The metric describes a
topological black hole for each $k$, with a unique event horizon located at
\begin{equation}
r_{+}=(2\mu G_{k}l^{2k})^{\frac{1}{d-1}}\;,  \label{Gammazero}
\end{equation}
for any positive $\mu $. Considering that at least one of the transverse directions must be
compact, this solution can be cast as a black ($d-3$)-brane\footnote{This solution can also be
interpreted as a ($d-2$)-brane with a least one spatial direction wrapped up.}, whose worldsheet
could be further wrapped. The vacuum configuration ($\mu =0$) is given by the locally AdS manifold,
described by the metric
\begin{equation}
ds^{2}=-\frac{r^{2}}{l^{2}}dt^{2}+\frac{l^{2}}{r^{2}}dr^{2}+r^{2}d\sigma
_{0}^{2}\;,  \label{vacuumgamma}
\end{equation}
which has no analogous in the vanishing cosmological constant limit.

${\bf \bullet \;\gamma =-1}$:

In this case, the set of metrics in Eq. (\ref{BHGeneral}) describe topological black holes which
possess a different behavior, depending on whether $k$ is odd or even. For all values of $k$, the
exceptional case $\mu =0$ possesses a horizon at $r_{+}=l$, which could be a black hole depending
on the topology of the transverse section $\Sigma _{-1}$. Both subcases --odd and even $k$-- must
be distinguished.

$\circ ${\bf \ Generic Theories with Odd }$k${\bf :}

For theories with odd $k$, if $\mu \geq 0$ there is always a single horizon
of radius $r_{+}\geq l$. The causal structure is the same as that of the
case $\gamma =1$,$\,$discussed above.

It is noteworthy that black hole solutions with negative mass densities can also exist for odd
$k$. If $\mu _{c}<\mu <0$, with $\mu _{c}$ given by (\ref {criticalmu}), the singularity at $r=0$
is surrounded by two horizons, $ r_{-} $ and $r_{+}$, and the causal structure is analogous to
that of the Reissner-Nordstrom solution with negative cosmological constant. For the critical
value $\mu =\mu _{c}$ both horizons coalesce at

\begin{equation}
r_{c}=l\sqrt{\frac{d-2k-1}{d-1}},  \label{rc}
\end{equation}
which corresponds to the extremal solution. The critical radius $r_{c}$ is the smallest possible
size of the outer horizon for the black holes within this family.

$\circ ${\bf \ Generic Theories with Even }$k${\bf :}

For each even $k$ there are two families of solutions labeled by $\alpha
=\pm 1$ in (\ref{BHGeneral}) with positive mass.

The branch with $\alpha=1$, describes black holes for $\mu \geq 0$ with a single horizon at
$r_{+}\geq l$, and with the usual causal structure. The bound $\mu =0$ is saturated by the metric
(\ref{massless}).

The black hole solutions belonging to the branch with $\alpha =-1$, have an unusual mass range,
bounded above and below by $\mu _{c}\geq \mu \geq 0$, which in terms of the horizon radius means
$r_{c}\leq r_{+}\leq l$. The positive upper bound $\mu _{c}$ is given by (\ref{criticalmu}). For
$\mu _{c}>\mu >0$ the solutions have two horizons $r_{-}$ and $r_{+}$, however, unlike the
standard solutions, as the mass increases, $r_{+}$ decreases. At first sight this might seem to
contradict the second law of thermodynamics, but this is not the case. The configuration $\mu=0$
will be excluded on thermodynamic grounds, as will be shown in section IV.

The extreme case, $\mu =\mu _{c}$ corresponds to the limit in which the horizons merge at $r_{c}$
given by (\ref{rc}), which is the smallest possible radius also in this case.

It is worth noting that if one considers a fixed mass parameter in the range $\mu _{c}\geq \mu
>0$, there exist two different topological black hole solutions, corresponding to the $\alpha =+1$
and $\alpha =-1$ branches, whose horizon radii are larger and smaller than $l$, respectively.

\subsection{Chern-Simons theories: $d=2k+1$}

In these theories, the functions $\alpha \left( \frac{2G_{k}\mu}{r^{d-2k-1}}\right)^{1/k}$
degenerate into horizontal straight lines and therefore $f^{2}(r)$ in Eq. (\ref{Delta}) possesses
only a simple zero at
\[
r_{+}=l\sqrt{\alpha \left( 2G_{k}\mu \right) ^{2/(d-1)}-\gamma }.
\]
This means that these solutions are black holes with a unique event horizon. Again, it is
necessary to distinguish the theories with even and odd $k$, corresponding to dimensions $d=4n+1$
and $d=4n-1$ respectively. The following table shows the allowed range of $\mu $ for which black
holes exist ($r_{+}\geq 0$) \vspace{0.5cm}
\begin{center}
{\large Chern-Simons theories}
\end{center}

\begin{equation}
\begin{array}{|l|c|c|c|}
\hline
& \gamma =1 & \gamma =0 & \gamma =-1 \\ \hline
&  &  &  \\
\begin{array}{l}
\text{odd }k\text{ } \\
(\alpha =1)
\end{array}
& \mu \geq \frac{1}{2G_{k}} & \mu > 0 & \mu \geq -{\frac{1}{2G_{k}}} \\ &  &  &  \\ \hline
\text{even }k & \mu \geq  \frac{1}{2G_{k}} & \mu > 0 &
\begin{array}{ll}
\alpha =1 & :\mu \geq 0 \\
&  \\
\alpha =-1 & :{\frac{1}{2G_{k}}}\geq \mu \geq 0
\end{array}
\\ \hline
\end{array}
\label{CHMassTable}
\end{equation}

As in the previous case, black holes with different values of $\gamma $ are
analyzed separately.

${\bf \bullet \;\gamma =1}$:

The spherical black holes were discussed in Ref. \cite{Banados:1994ur}, and in further detail in
\cite{Crisostomo:2000bb}. As seen in Eq. (\ref{mass}), the lower bound $\mu =\frac{1}{2G_{k}}$
corresponds to the zero mass black hole ($M=0$), which is separated by a mass gap from AdS
spacetime ($M=- \frac{1}{2G_{k}}$). These black holes have a common causal structure with the
($2+1$)-dimensional solution \cite{Banados:1992wn}.

${\bf \bullet \;\gamma =0}$:

As for the generic theories ($d-2k\neq 1$), the locally flat transverse section $\Sigma _{0}$, is
assumed to be orientable with at least one compact direction. In that theory, the metric
(\ref{BHGeneral}) describes a black ($d-3$)-brane, whose horizon is located at $r_{+}=l(2G_{k}\mu
)^{1/(d-1)}$, as is obtained from (\ref{Gammazero}) for $d-2k=1$. Unlike the $\gamma =1$ case, the
black brane vacuum ($\mu =0$) corresponds to the same metric as in the generic case given by
(\ref{vacuumgamma}), and there is no energy gap.

${\bf \bullet \;\gamma =-1}$:

As in the generic theories, solutions (\ref{BHGeneral}) describe topological black holes for the
range of masses included in table (\ref{CHMassTable}), except that for $\mu =0$, the metric
(\ref{massless}) may or may not be a black hole, depending on the topology of $\Sigma _{-1}$.
Unlike the generic theory, these family of topological black holes possess a single event horizon
$r_{+}$ even for negative values of $\mu$.

The minimum size of this kind of black holes is $r_{c}=0$, as can be seen from Eq. (\ref{rc}),
whose critical mass parameter is given by
\begin{equation}\label{mucs}
\mu _{c}={\frac{(-1)^{k}}{2G_{k}}}.
\end{equation}

As in the generic case, the massive solutions of Chern-Simons ({\bf CS}) theories have different
features, for odd and even $k$.

$\circ $ {\bf CS theories with Odd }$k${\bf \ (}$d=4n-1${\bf ):}

The solution (\ref{BHGeneral}) describes black holes with a single event horizon for $\mu \geq \mu
_{c}=-\frac{1}{2G_{k}}$, and naked singularities otherwise.

$\circ ${\bf \ CS theories with Even }$k${\bf \ (}$d=4n+1${\bf ):}

Two families of solutions with positive mass labeled by $\alpha =\pm 1$ are obtained.

The branch with $\alpha =+1$, describes black holes with $\mu \geq 0$ and $ r_{+}\geq l$, where
the bound $\mu =0$ is saturated by (\ref{massless}).

The mass range of the black holes with $\alpha =-1$ is bounded above and below by $0\leq \mu \leq
\mu _{c}=\frac{1}{2G_{k}}$, which in terms of the horizon radius means $l\geq r_{+}\geq 0$. Note
that for $\alpha =-1$, the mass is a decreasing function of $r_{+}$.

If the mass parameter is in the range $\mu _{c}\geq \mu >0$, two inequivalent topological black
hole solutions are found, corresponding to the branches $\alpha =\pm 1$, as in the generic theory.

Note that the static $2+1$ black hole is obtained from Eq. (\ref{BHGeneral}) for $\gamma =1$ as
well as for $\gamma = 0$, because in three dimensions the transverse section degenerates to
$S^{1}$.

\subsection{Vanishing cosmological constant limit}

The full set of topological black hole metrics discussed here approach asymptotically a locally
AdS spacetime with radius $l$, whose curvature at the boundary satisfies $R^{ab}\rightarrow
-l^{-2}e^{a}e^{b}$. Hence, the asymptotically flat limit is obtained for $l\rightarrow \infty $,
instead of taking the vanishing limit of the volume term $(c_{0}^{k}\rightarrow 0)$. The
vanishing cosmological constant limit of the solutions in Eq. (\ref {BHGeneral}) coincides with
the solutions of the $l\rightarrow \infty $ limit of the action $I_{k}$, or equivalently, taking
the same limit in the field equations (\ref{kEinstein}) and (\ref{kTorsion}), which amounts to
replacing $\bar{R}^{ab}$\ by $R^{ab}$ \cite{Crisostomo:2000bb}.

The asymptotically flat limit of (\ref{BHGeneral}) is given by
\begin{eqnarray}
ds^{2} &=&-\left( \gamma -\alpha \left( \frac{2G_{k}\mu }{r^{d-2k-1}}\right)
^{1/k}\right) dt^{2} +  \nonumber \\
&& \frac{dr^{2}}{\gamma -\alpha \left( \frac{2G_{k}\mu }{r^{d-2k-1}} \right)^{1/k}} + r^{2}
d\sigma_{\gamma}^{2}.  \label{BHSL}
\end{eqnarray}
Hence, in case of vanishing $\Lambda$, these metrics describe black holes only for the spherically
symmetric solutions in the non-CS case ($\gamma =1$, and $d-2k-1\neq 0$), with an event horizon
located at $ r_{+}=(2G_{k}M)^{1/(d-2k-1)}$.

Some of the topological black holes discussed here have been previously reported elsewhere. The
case of Einstein-Hilbert action --corresponding to $ k=1$ in our analysis--, was extensively
studied in \cite{Vanzo:1997gw} and \cite{Brill:1997mf}. The topological black holes corresponding
to $k=1$, possess a geometry resembling just those found for the actions $I_{k}$ in Eq.
(\ref{BHGeneral}) with odd $k$; in fact, they possess the same causal structure. However, as is
shown below, the thermodynamic behavior corresponding to the Einstein-Hilbert case differs from
the other odd $k$ theories.

The theories with $k=[\frac{d-1}{2}]$ were studied in \cite{Cai:1998vy} for odd $k$, which
correspond to Chern-Simons and Born-Infeld theories in dimensions $d=4n-1$, $4n$ respectively.

\section{Thermodynamics}

\subsection{Temperature and specific heat}

The black hole temperature is defined in the standard way as $\beta =\frac{1}{\kappa _{B}T}$,
where $\kappa _{B}$ is the Boltzmann constant, and the period $\beta =4\pi \left( \left.
\frac{df^{2}}{dr}\right| _{r_{+}}\right) ^{-1}$ is found by demanding regularity of the Euclidean
solution at the horizon. Thus,
\begin{equation}
T=\frac{(d-1)}{4\pi \kappa _{B}l^{2}}\frac{r_{+}^{2}+\gamma r_{c}^{2}}{ kr_{+} },
\label{Temperature}
\end{equation}
where $r_{c}$ is the critial radius defined in Eq. (\ref{rc}). Note that for CS theories the
temperature has the universal expression
\begin{equation}
T_{\text{CS}}= \frac{r_{+}}{2\pi\kappa_{B}l^2},
\end{equation}
which does not depend on $d$ or $\gamma$. For generic theories with $\gamma=0$, the temperature is
also a linear function of $r_{+}$, and this result is approximated for $r_{+}\gg l$ in all the
other cases.
\begin{figure}[tbm]
\begin{center}
  \leavevmode
  \epsfxsize=3 in
\epsfbox{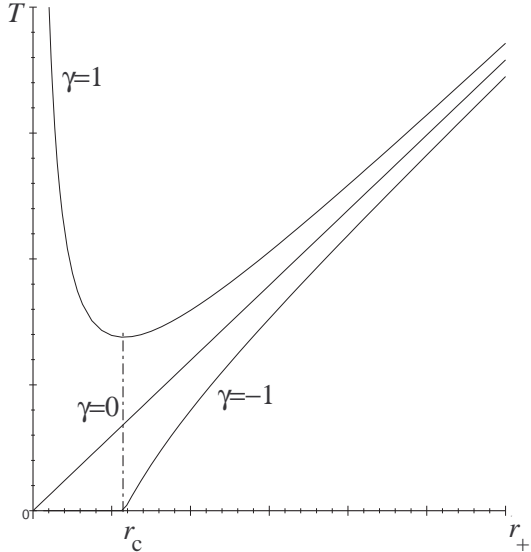} \caption{The temperature as a function of $r_{+}$ is depicted for generic
theories with $\gamma=0,\pm 1$. For $\gamma=1$ the temperature has a minimum at $r_{+}=r_{c}$.
When $\gamma=0$ the temperature is a linear function of $r_{+}$. For $\gamma=-1$ the temperature
is an increasing function of $r_{+}$ that vanishes at $r_{+}=r_{c}$. For $r_{+}\gg l$ the
temperature grows linearly with $r_{+}$ for all cases. For CS theories the three curves are
replaced by the $\gamma=0$ straight line with a universal slope.}
\end{center}
\end{figure}
The specific heat $C=\frac{dM}{dT}$ is given by
\begin{equation}
C=k\frac{2\pi \kappa _{B}}{G_{k}}\frac{\Omega _{d-2}}{\Sigma _{d-2}}
r_{+}^{d-2k}\left( \gamma +\frac{r_{+}^{2}}{l^{2}}\right) ^{k-1}\left[ \frac{
r_{+}^{2}+\gamma r_{c}^{2}}{r_{+}^{2}-\gamma r_{c}^{2}}\right] ,
\label{HeatCapacity}
\end{equation}
which for $r_{+}>>l$ grows like $C\sim r_{+}^{d-2}$. Combining formulas (\ref{Temperature}) and
(\ref{HeatCapacity}) with the mass parameter $\mu (r_{+})$ defined in Eq. (\ref{mu}), it is
possible to investigate whether these topological black holes can reach thermal equilibrium with a
heat bath at temperature $T_{B}$.

\subsection{Thermal equilibrium}\label{ThermalEquilibrium}

$\bullet \;\gamma =1:$

The thermodynamic equilibrium of spherically symmetric black holes ($\gamma =1$) was discussed in
Ref. \cite{Crisostomo:2000bb}. In this theory, for $ d-2k-1\neq 0$, the temperature
(\ref{Temperature}) has a minimum $T_{c}= \frac{\sqrt{\left(d-2k-1\right) \left( d-1\right)
}}{2\pi \kappa _{B}kl}$ at $r_{+}=r_{c}=l \sqrt{\frac{d-2k-1}{d-1}}$. The specific heat (\ref
{HeatCapacity}) is positive for $r_{+}>r_{c}$, and has the opposite sign for $r_{+}<r_{c}$; and
near the critical radius behaves as $C\sim (r_{+}-r_{c})^{-1}$, signaling the existence of a phase
transition. Two generic situations may occur.

{\bf (i)} $T_{B}>T_{c}$: In this case there are two possible equilibrium states of radii $r_{u}$
(unstable) and $r_{s}$ (locally stable), with $ r_{u}<r_{c}<r_{s}$. Thus, if the initial state has
$r_{+}<r_{u}$, the black hole cannot reach the equilibrium because it evaporates until its final
stage. Otherwise, for $r_{+}>r_{u}$, the black hole evolves towards an equilibrium
configuration\footnote{ Curiously, the minimum size for which a spherical black hole can be at
equilibrium with a heath bath ($r_{c}$) corresponds to the smallest size of a topological black
hole with hyperbolical transverse section.} at $ r_{+}=r_{s}$.

{\bf (ii)} $T_{B}<T_{c}$: Under this assumption, a black hole cannot reach a
stable equilibrium state and is doomed to evaporate.

In the special case of $d-2k=1$ (CS), the specific heat (\ref{HeatCapacity}) is always positive,
hence the equilibrium configuration is always reached, independently of the initial black hole
state and for any finite temperature $T_{B}$.

$\bullet \;\gamma =0:$

In case of black holes with a locally flat transverse section ($\gamma =0$), the temperature
(\ref{Temperature}) grows linearly with $r_{+}$
\begin{equation}
T=\frac{1}{k}\frac{(d-1)}{4\pi \kappa _{B}}\frac{r_{+}}{l^{2}}\;,  \label{T0}
\end{equation}
and the specific heat is
\[
C=k\frac{2\pi \kappa _{B}}{G_{k}l^{2k-2}}\frac{\Omega _{d-2}}{\Sigma _{d-2}}
r_{+}^{d-2}\;,
\]
which implies that, independently of the initial black hole state, thermal equilibrium at some
$r_{+}=r_{s}$, is always reached for any finite temperature of the heat bath $T_{B}$.

$\bullet \;\gamma =-1:$

For all theories labeled with different $k$, the temperature of the
topological black holes with hyperbolic transverse sections is a
monotonically increasing function of $r_{+}$ which vanishes at the smallest
possible size for a black hole, $r_{+}=r_{c}$. This is consistent with the
fact that for the extremal solution, $r_{+}=r_{c}$, the Euclidean $r-t$
plane has the topology of a cylinder and hence $\beta $ is arbitrary.

The massless topological black hole in Eq. (\ref{massless}) has a horizon at
$r_{+}=l$ and a universal temperature given by
\begin{equation}
T_{l}=\frac{1}{2\pi \kappa _{B}l}.  \label{Tl}
\end{equation}

The specific heat (\ref{HeatCapacity}) has a simple zero at $r_{+}=r_{c}$ and a zero of order
$k-1$ at $r_{+}=l$. This second root is a local minimum for odd $k$ and a saddle point\footnote{
Except for $k=2$, in which case $C$ has a simple zero at $r_{+}=l$.} for even $k$. Thus, the
approach to equilibrium depends on the parity of the integer $k$.
\begin{figure}[tbm]
\begin{center}
  \leavevmode
  \epsfxsize=2.5 in
\epsfbox{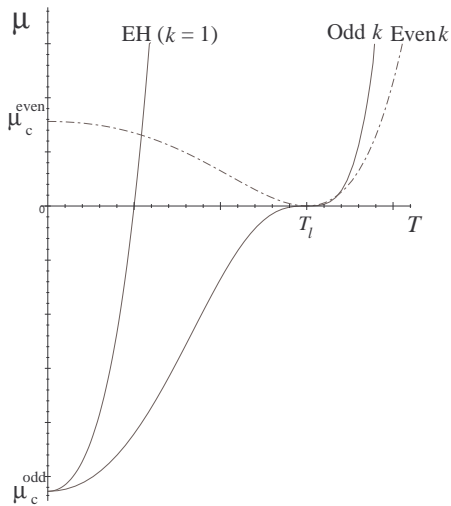} \caption{The mass parameter ($\mu$) as function of the temperature is depicted
for $\gamma=-1$ solutions. For even $k$, $\mu$ has a local maximum at $T=0$ and an absolute
minimum at $T_{l}$. For odd $k$, the mass parameter has an absolute minimum at $T=0$ and an
inflexion point at $T_{l}$ for $k\neq 1$. The specific heat vanishes at these critical points. For
even $k$ the specific heat is negative for $T<T_{l}$. For odd $k\neq 1$, the inflexion point at
$T=T_{l}$ signals the existence of a ``volatile point''. For the EH case ($k=1$) there is not such
volatile state, since the specific heat has an absolute minimum at $T=0$.}
\end{center}
\end{figure}

\subsubsection{Theories with odd $k$:}

For $k\neq 1$, as depicted in Figs. 2 and 3, the temperature is a strictly increasing function of
$\mu $ and the specific heat is a nonnegative for the allowed range of $r_{+}$, for $k\neq 1$. This
implies that equilibrium with a heat bath at temperature $T_{B}$ is always reached for any initial
black hole state. Moreover, since the specific heat vanishes for the local minimum at $r_{+}=l$ ,
the topological black hole behaves as a ``volatile'' system near the massless configuration, as
there is a sudden increase in temperature with an infinitesimal increase in $\mu $.

For the EH action ($k=1$) the specific heat neither vanishes nor has a minimum at $r_{+}=l$, as
seen in Fig. 3. This means that equilibrium with a heat bath is also attained for any initial
configuration, and there is no ``volatility point'' at all.

\subsubsection{Theories with even $k$:}

As shown in Fig. 3, the topological black holes in all theories with even $k$ have positive
specific heat for $r_{+}>l$ -corresponding to the branch with $ \alpha =+1$-, while $C<0$ for
$r_{c}<r_{+}<l$ (for $\alpha =-1$ ). The zero of the specific heat at $r_{+}=l$, which corresponds
to the massless topological black hole in Eq. (\ref{massless}) at temperature $T_{l}$, gives rise
to two different scenarios depending whether the heat bath temperature is above or below $T_{l}$.

{\bf (i)} $T_{B}<T_{l}$: If the initial state of the topological black hole is at a temperature
above $T_{B}$, both branches ($\alpha =\pm 1$) reduce their masses, reaching a stable vacuum
configuration out of thermal equilibrium at temperature $T_{l}$ and zero mass. On the contrary, if
the initial state is at $T<T_{B}$ (which can only occur for the branch $\alpha =-1$), the black
hole increases its mass tending towards the extreme state with $T=0$ and $\mu =\mu _{c}$. Hence,
the configuration at thermal equilibrium ($T=T_{B}$) is unstable.

{\bf (ii) }$T_{B}>T_{l}$: If the initial state of the topological black hole belongs to the upper
branch ($\alpha =+1$ and $T>T_{l}$), the equilibrium with the heat bath is always attained, in
agreement with the positive specific heat of this branch. Conversely, a black hole in the lower
branch will move away, reducing its temperature and increasing its mass, towards the extreme
configuration. Note that now the vacuum configuration $\mu =0$ is unstable.

\subsubsection{Thermodynamics and topology fixing for $\gamma =-1$}

As mentioned in Sec. \ref{TopologicalSolutions} for the $\gamma =-1$ case, the massless state
(\ref{massless}) could be construed as a black hole or not, depending on the topology of the
transverse section. The above thermodynamic analysis shows that the state $\mu =0$ admits a
standard black hole interpretation for odd $ k$ but not for even $k$.

To see this, consider a nearly massless black hole in vacuum (in a heat bath at zero temperature):
in a theory with even $k$ both branches ($\alpha =\pm 1 $) approach a final state at $\mu =0$ and
$T=T_{l}$ which cannot loose energy further as that would make the metric complex. In this sense,
this final state cannot be interpreted as a standard black hole. On the other hand, for theories
with odd $k$ the $\mu =0$ configuration behaves as a volatile state which radiates violently,
decaying into a negative mass black hole.

Thus, the thermodynamics provides a criterion to restrict the topology of the transverse section
$\Sigma _{-1}$: for odd $k$, the transverse section must have a topology such that the geometry
for $\mu =0$ {\em is }a black hole; on the contrary, for even $k$, the topology of the transverse
section must be chosen so that the massless solution {\em is not }a black hole.

\subsection{Entropy}

\label{SectionEntropy}

An analytic expression for Entropy as a function of the horizon radius $r_{+}$, can be obtained in
the semiclassical approximation from the Euclidean version of the action $I_{k}$ (See Appendix).
Alternatively, the entropy can be obtained from the first law of thermodynamics, $dM = TdS$, as
\begin{equation}
\frac{dS_{k}}{dr_{+}}=\frac{\Sigma _{d-2}}{T\Omega _{d-2}}\frac{d\mu }{dr_{+} },  \label{DS}
\end{equation}
which, upon substitution of Eqs. (\ref{mu}) and (\ref{Temperature}), yields
\begin{equation}
\frac{dS_{k}}{dr_{+}}=k\frac{2\pi \kappa _{B}\Sigma _{d-2}}{\Omega
_{d-2}G_{k}}\,r_{+}^{(d-2k-1)}\left( \gamma +\frac{r_{+}^{2}}{l^{2}}\right)
^{k-1}.  \label{DEntropy}
\end{equation}
This expression shows that the entropy is a monotonically increasing function of $r_{+}$ for all
cases, except if $k$ is even and $\gamma =-1$, in which case it is decreasing in the range
$r_{c}<r<l$.

For $\gamma =1,0$ the entropy $S_{k}(r_{+})$ is given by
\begin{equation}
S_{k}= \kappa _{B}\frac{2k\pi\Sigma _{d-2}}{\Omega _{d-2}G_{k}}\int_{r_{
\text{min}}}^{r_{+}}\,r^{(d-2k-1)}\left( \gamma +\frac{r^{2}}{l^{2}}\right) ^{k-1}dr\;,
\label{Entropy}
\end{equation}
for which the lower integration limit is chosen as $r_{\text{min}}=0,$ so that the vacuum
$(r_{+}=0)$ has vanishing entropy. For $\gamma =-1$ , expression (\ref{Entropy}) is valid also for
theories with odd $k$, provided $r_{+}>r_{c}$.

In the exceptional case (even{\bf \ $k$, }$\gamma =-1$), Eq.(\ref{DEntropy}) implies that the
entropy attains an absolute minimum at the vacuum configuration ($r_{+}=l$). Hence, the lower
integration limit is naturally chosen as $r_{\text{min}}=l$ in order to have nonnegative
entropy\footnote{ Note that in this case the entropy vanishes for $r_{+}=l$, where the temperature
is nonzero ($T=(2\pi \kappa _{B}l)^{-1}$). This situation is not completely new, as it is found
for instance in stringy dilatonic black holes \cite{Holzhey:1992bx}. These configurations are
physically acceptable provided the black hole has a mass gap, a condition which is met by the
solutions presented here.}. Superficially, the fact that in the range $r_{c}<r_{+}<l$, the entropy
is a decreasing function of $r_{+}$, would seem to violate the second law of thermodynamics.
However, this range of $r_{+}$ corresponds to the branch with $\alpha =-1$, for which the mass is
also a decreasing function of $r_{+}$ and hence $\partial S/{\partial M}=1/T$, which is a positive
quantity, as shown in Fig. 2.

All solutions with $\gamma =0$ obey an ``area law'' for all $k$:
\begin{equation}
S_{k}=\kappa _{B}\frac{2\pi k}{(d-2)\Omega _{d-2}G_{k}l^{2(k-1)}}A\;, \label{S(A)}
\end{equation}
which in standard units is
\[
S_{k}=k\frac{G}{G_{k}l^{2(k-1)}}S_{EH},
\]
where the Einstein Hilbert entropy reads $S_{EH}=\frac{\kappa_{B}}{\tilde{G}}\frac{A}{4}$, and
$A=\Sigma _{d-2}r_{+}^{d-2}$ is the horizon area. It is important to note that for $ \gamma \pm 1$
the entropy (\ref{Entropy}) approaches the area law (\ref{S(A)}) in the limit $r_{+}\gg l$.

\subsection{Canonical Ensemble}

For spherically symmetric black holes in the four-dimensional Einstein theory, it was shown in
Ref. \cite{Hawking:1983dh} that the canonical ensemble is well defined provided a negative
cosmological constant is included. Those arguments can be extended to higher dimensions and for
theories described by the action (\ref{Ik}) \cite{Crisostomo:2000bb}. It can be similarly shown
that the canonical ensemble is also well defined for the whole set of topological black hole
solutions considered here.

In the present case the partition function is given by
\begin{equation}
Z_{\gamma }^{k}(\beta )=\sum_{\alpha }\int \rho _{\alpha }(M)e^{-\beta M}dM,
\label{Z(M)}
\end{equation}
where the sum extends over $\alpha =\pm 1$ only for even $k$\ and $\gamma =-1 $, while for all
other cases $\alpha =1$ only. Integrating in $r_{+}$, this expression reads
\begin{equation}
Z_{\gamma }^{k}(\beta )=\int_{r_{c}}^{\infty }\rho (r_{+})e^{-\beta M}\left|
\frac{\partial M}{\partial r_{+}}\right| dr_{+},  \label{Z(r)}
\end{equation}
for all cases. As the density of states is given by $\rho (r_{+})=\exp (S_{k}/\kappa _{B})$, the
convergence of (\ref{Z(r)}) depends on the behavior of $S_{k}$ and $M$ for $r_{+}\gg l$. Combining
Eqs. (\ref{mu}) and (\ref{S(A)}), the integral (\ref{Z(r)}) can be seen to converge for all $k$ and
$\gamma $. In fact, the integrand of (\ref{Z(r)}) takes the convergent form $\exp [-\beta M +
aM^\frac{{d-2}}{d-1}]$\ ($a>0$) for $\Lambda =-l^{2}\neq 0$, whereas for $\Lambda =0$, it behaves
as $\exp [-\beta M +a M^\frac{{d-2k}}{d-2k-1}]$ and diverges for all $k$.

\subsection{Connection with thermal CFT's.}

In the context of the Maldacena's AdS/CFT duality conjecture (see e.g. \cite {Aharony:1999ti} and
references therein), a $d$-dimensional Euclidean gravity theory with asymptotic AdS behavior can
be described by a suitable thermal conformal field theory on its boundary \cite{Witten:1998zw}.
The fact that the actions $I_{k}$ --defined in Eq.(\ref{Ik})-- describes up to $\left[
\frac{d-1}{2}\right] $ inequivalent gravity theories in the bulk, would imply the existence of an
equal number of different ($d-1$)-dimensional dual CFT's at the boundary. The asymptotic behavior
of these gravity theories, which can be read from the metrics (\ref{BHGeneral}), should be
reconstructed from this set of CFT's at the boundary through the UV/IR relation. This relation
states that different radial positions are mapped to different field theory scales, in such a way
that the infrared effects in the bulk correspond to ultraviolet effects on the theory at the
boundary \cite{Susskind:1998dq}.

In particular, certain type of CFT renormalization group equation can be generated by the action
of the radial Hamiltonian constraint in the bulk \cite{Balasubramanian:2000wv}. It is expected
that deviations from the proposed renormalization group flow should result from modifying the
Hamiltonian constraints by the inclusion of higher curvature terms in the action. Thus, the set of
actions given by $I_{k}^{d}$ enhances the repertoire of theories which could provide a concrete
holographic interpretation of gravity.

The different asymptotic regions of the black holes discussed here provide inequivalent background
spacetimes where the corresponding dual CFT's are realized \cite{Birmingham:1998nr,Emparan:1999pm}.
Thus, CFT's defined on $S^{1}\times {\mathbb{R}}^{d-2}$, $S^{1}\times S^{d-2}$ or $S^{1}\times
H^{d-2}$ are connected with black holes in the bulk for $\gamma=0 $, $1$, or $-1$, respectively.

Some insight about the correspondence can be gained by looking at the thermodynamic quantities in
the simplest case corresponding to a thermal CFT on a flat background, that is, on $S^{1}\times
{\mathbb{R}}^{d-2}$. In this case, conformal invariance is sufficient to fix the energy to be of
the form

\begin{equation}
E_{CFT}=\sigma _{SB}V_{CFT}\beta _{CFT}^{1-d}\;,  \label{ECFT}
\end{equation}
where $\sigma _{SB}$ is the Stefan-Boltzmann constant. Using the first law of thermodynamics, the
entropy can be written as\footnote{ The form of $S$ can also be inferred demanding the entropy to
be extensive and conformally invariant.}
\begin{equation}
S_{CFT}=\frac{d-1}{d-2}\sigma _{SB}^{\frac{1}{d-1}}V_{CFT}\left( \frac{ E_{CFT}}{V_{CFT}}\right)
^{\frac{d-2}{d-1}}\;,  \label{SCFT}
\end{equation}

The precise expression for $\sigma _{SB}$ is determined by the dynamical structure of the specific
CFT considered, and is a growing function of the number of degrees of freedom. In terms of the
AdS/CFT correspondence, this CFT can be viewed as defined on the boundary of the Euclidean space
for the metric (\ref{BHGeneral}) with $\gamma =0$ at a large fixed radius $r_{0}\gg r_{+}$. Hence,
$V_{CFT}$ corresponds to the volume of the transverse section, given by

\begin{equation}
V_{CFT}=\Sigma _{d-2}r_{0}^{d-2}.  \label{VCFT}
\end{equation}
Correspondingly, the temperature at $r_{0}$ is given by the red-shifted black hole temperature as
\[
T_{CFT}=\frac{l}{r_{0}}T\;,
\]
where $T$ is given by Eq. (\ref{T0}). Consequently, the energy $E_{CFT}$ in Eq. (\ref{ECFT})
corresponds to the red-shifted black hole mass,
\begin{equation}
E_{CFT}=\frac{l}{r_{0}}M,  \label{RedshiftedE}
\end{equation}
provided the Stefan-Boltzmann constant is given by
\begin{equation}
\sigma _{SB}=\frac{1}{2\Omega _{d-2}}\left( \frac{4\pi k}{d-1}\right) ^{d-1}
\frac{l^{d-2k}}{G_{k}}\;.  \label{Steffan-Boltzmann}
\end{equation}
Plugging expressions (\ref{ECFT}), (\ref{VCFT}) and (\ref{Steffan-Boltzmann} ) into Eq.
(\ref{SCFT}), allows expressing $S_{CFT}$ in terms of the black hole mass density,
\begin{equation}
S_{CFT}=k\left( \frac{l^{2(d-k-1)}}{G_{k}}\right) ^{\frac{1}{d-1}}\frac{2\pi \kappa
_{B}}{(d-2)}\frac{\Sigma _{d-2}}{\Omega _{d-2}}(2\mu )^{\frac{d-2}{d-1 }},  \label{Scft}
\end{equation}
which precisely matches the black hole entropy $S_{k}$ for $\gamma =0$ in Eq. (\ref{S(A)}).

It is worth noting that the entropy grows linearly with $k$, the highest power of curvature in the
action. Moreover, Eq. (\ref{Steffan-Boltzmann}) relates the integer $k$ to the number of degrees
of freedom of the corresponding CFT at strong coupling. For the case of standard five-dimensional
supergravity ($k=1$), which is conjectured to be dual to four dimensional SYM with ${\cal N}=4$ at
large $N$, the entropy relation (\ref{Scft}) is reproduced up to a numerical factor
\cite{Witten:1998zw}.

Finally, note that the entropy matching between black holes with $\gamma =0$ and CFT's on a flat
background $S^{1}\times {\Bbb R}^{d-2}$ is exact for all values of $\beta $, but it is not
necessarily so for CFT's defined on $ S^{1}\times S^{d-2}$ and $S^{1}\times H^{d-2}$, i.e.,$\gamma
=1$ and $-1$ respectively. Although the exact expression for the entropy of a CFT on a generic
curved background is unknown, an approximate result can be established for $\gamma =\pm 1$ in the
limit $\beta \rightarrow 0$. In fact, for $\gamma =\pm 1$,\ the curvature of the transverse section
is $\pm 1/r_{0}^{2}$\ and, by conformal invariance, the entropy should be a function of $\beta
/r_{0}$\ only. Hence, the large $r_{0}$\ limit is equivalent to $ \beta \rightarrow 0$\ and
therefore the high temperature limit is reproduced if the horizon radius is very large \footnote{
In generic theories (non CS) theories with $\gamma =1$, the limit $\beta \rightarrow 0$ can be
obtained for $r_{+}<<l$. This branch, however, is thermodynamically unstable and this fact could be
interpreted as a confined phase in the CFT \cite{Witten:1998zw}.} ($r_{+}>>l$). Thus it is
concluded that the entropy of a CFT and that of a black hole approach the same expression given by
(\ref{S(A)}) in the high temperature limit, provided $\sigma_{SB}$ is chosen as in Eq.
(\ref{Steffan-Boltzmann}).

\section{Summary and comments}

Static black hole-like geometries, possessing topologically nontrivial AdS asymptotics have been
found as solutions of a family of gravity theories which admit a unique global AdS vacuum. These
theories and their corresponding solutions are classified by an integer $k$, which is the highest
power of curvature in the Lagrangian. These solutions are further labeled by the constant $\gamma
=\pm 1,0$, representing the curvature of the transverse section.

{\bf Locally spherical transverse section:} The case $\gamma=1$ leads to a natural splitting
between generic and CS theories ($d-2k=1$). In the first case, the causal and thermodynamic
properties resemble those of the Schwarzschild-AdS black hole. In the CS case, black holes behave
like the $2+1$ solution.

{\bf Locally flat transverse section:} The case $\gamma=0$ corresponds to (un-)wrapped black
branes, for all values of $d$, $k$\ and $M>0$ exhibiting the same causal structure as a
Schwarzschild-AdS black hole, but whose thermodynamic properties are analogous to those of a $2+1$
dimensional black hole. Therefore, they possess a single event horizon, their temperature is a
linear function of $r_{+}$, and hence they reach thermal equilibrium with a heat bath at any
temperature. Moreover, the entropy follows an area law, $S_{k}=\kappa_{B}\frac{2\pi k}{(d-2)\Omega
_{d-2}G_{k}l^{2(k-1)}}A$.

{\bf Locally hyperbolic transverse section:} The case $\gamma=-1$ naturally leads to a splitting
between theories with even and odd $k$ which are treated separately. In addition, the
Einstein-Hilbert action ($k=1$) and the CS cases exhibit a special behavior.

{\bf a)Odd $k$.} In this case, solutions with non-negative mass have a single horizon radius larger
than $l$ and their causal structure is analogous to that of the $\gamma =1$ case discussed above.
For $M<0$, in the generic case there are two horizons with the same causal structure as the
Reissner-Nordstrom AdS black hole, but where $r_{+}$ and $r_{-}$ cannot be independently adjusted
because they are functions of a single parameter ($\mu$). The extremal case corresponds to the
lower bound for both mass and $r_+$ ($M\geq M_{c}<0$ and $r_+ \geq r_{c}$). The temperature is a
strictly increasing function of the mass and hence the specific heat is non-negative for the entire
physical range ($r_{+}\geq r_{c}$). This means that equilibrium with a heat bath can always be
reached. The specific heat vanishes at $r_{+}=l$, signaling the existence of ``volatile''
configurations near the massless state. The Einstein theory ($k=1$) is singled out in this
respect, since its specific heat neither vanishes nor has a minimum at the massless configuration
and therefore exhibits no volatile behavior.

{\bf b) Even $k$.} These theories possess an interesting set of black hole solutions with
hyperbolic transverse section. In this case, there exist two independent branches for a given
mass: the branch with $\alpha =1$ describes single horizon black holes with $r_{+}\geq l$. They
have non-negative mass and the usual causal structure. The other branch ($\alpha =-1$), has a
non-standard mass range $\mu _{c}\geq \mu > 0$, and the corresponding range of horizon radius is
$r_{c}\leq r_{+}< l$. On the other hand, solutions belonging to this latter branch present two
horizons and curiously, $r_{+}$ is a decreasing function of the mass, unlike the standard black
holes. The extreme case corresponds to the smallest possible size of the horizon radius ($r_{c}$),
which has the largest possible mass ($ \mu =\mu _{c}$).

The following remarks on the thermodynamics are in order.

$\bullet$ Topological black holes with hyperbolic transverse section and even $k$ can reach thermal
equilibrium only if the temperature of the bath is higher than that of the massless $(r_{+}=l)$
configuration ($T_{B}>T_{l}$), and if the initial state of the black hole belongs to the upper
branch ($\alpha =+1$ and $T>T_{l}$). Otherwise, the fate of the black hole is to approach either
the vacuum ($\mu=0$), or the extremal configuration ($\mu=\mu_c$), as discussed in section
\ref{ThermalEquilibrium}.

$\bullet$ It is remarkable that for $\gamma =-1$, thermodynamics restricts the topology of the
transverse section $\Sigma_{-1}$: for even $k$ it must be such that the configuration $\mu =0$ is
not a black hole, whereas for odd $k$, the massless configuration must be a black hole, which for
instance can be obtain through suitable identification in the covering space of the transverse
section.

$\bullet$ For CS theories the temperature has a universal linear dependence on $r_{+}$ for all $d$
and $\gamma$.

$\bullet$ Solutions with $\gamma =-1$, of CS theories always possess a single horizon and have
qualitative thermodynamic behavior for even and odd $k$.

$\bullet$ The canonical ensemble is well defined for all values of the parameters $d,k,\mu$ and
$\gamma$, provided a negative cosmological constant is present. Otherwise, the partition function
diverges.

$\bullet$ In the vanishing cosmological constant limit, only the spherically symmetric solutions
($\gamma =1$) with $d-2k\neq 1$ are black holes.

{\bf Holography:} The topological black hole solutions found here shed some light on holography in
the sense of the AdS/CFT correspondence. It has been shown that the black hole thermodynamics for
$\gamma=0$ can be described in terms of a CFT at the boundary, for all the theories considered
here. For $\gamma=\pm 1$ the matching occurs for $r_+ \gg l$. Thus, Einstein's theory is not the
only one which admits a holographic interpretation, but the whole set of gravitational theories
presented here do. The exact matching with a CFT is achieved provided the value of the
Steffan-Boltzmann constant is a fixed function of $d$ and $k$ given by Eq.
(\ref{Steffan-Boltzmann}). Since $\sigma_{SB} \sim k^{d-1}$, the number of degrees of freedom in
the CFT must increase with the power of the curvature in the bulk gravitational theory. Hence, the
AdS/CFT correspondence, in this sense, suggests the existence of $\left[ \frac{d-1}{2} \right]$
inequivalent ($d-1$)-dimensional dual CFT's, one for each action $I_{k}$, enlarging the options
for concrete holographic interpretation of gravity. In particular, in five dimensions there are two
gravitation actions within this family (EH and CS), which are candidates in equal footing to
realize the AdS/CFT correspondence.

One could speculate that for CS gravity, the asymptotic dynamics would be described by some higher
dimensional generalization of the WZW model (See, e.g., Refs.
\cite{Nair:1990aa,Banados:1996yj,Chandia:1998uf,GegenbergKunstatter}). Thus, WZW models may be
relevant to count the microstates responsible for the entropy of these black holes.

Topological black hole metrics in eleven dimensions with $k=5$ are also solutions of a
supergravity theory, described in terms of a CS action with gauge group $0Sp(32|1)$
\cite{Troncoso:1998va}. Furthermore, it can be shown that some of them admit Killing spinors
\cite{InProgress}. This claim might seem surprising as no local supersymmetric extension exists
for the EH action with cosmological constant in eleven dimensions \cite{Bautier:1997yp}.

\section{Acknowledgements}

The authors would like to thank Jorma Louko, Rodrigo Soto and Luciano Vanzo for enlightening
comments. This work was supported in part through grants 3990009 and 1990189 from FONDECYT, and
grant DI 51-A/99 (UNAB). The institutional support of I. Municipalidad de Las Condes, and a group
of Chilean companies (AFP Provida, CODELCO, Empresas CMPC, MASISA S.A. and Telef\'{o}nica del
Sur). CECS is a Millennium Science Institute.
\appendix

\section{Mass and Entropy from boundary terms}

\subsection{\label{HamiltonianCharges}Mass}

The aim of this appendix is to establish the relationship between the integration constant $\mu$,
appearing in the solutions (\ref{BHGeneral}), and the mass. In the Hamiltonian approach, the
gravitational action is
\begin{equation}
I_{T}=I_{G}+B,  \label{IHamiltonian}
\end{equation}
where $I_{G}$ is the canonical action in phase space,
\begin{equation}  \label{IMH}
I_{G}=\int d^{d}x(\pi ^{ij}\dot{g}_{ij}-N^{\bot }H_{\bot }-N^{i}H_{i}),
\label{IGH}
\end{equation}
and $B$ is a boundary term, which is required in order to guarantee that the action attain an
extremum on shell \cite{Regge:1974zd}. Here $H_{\mu }$ are the Hamiltonian generators of spacetime
diffeomorphisms.

Replacing the ansatz (\ref{ds}) into the action, allows to obtain a one-dimensional minisuperspace
model whose action,
\begin{equation}
I_{T}=\Delta t\frac{\Sigma _{d-2}}{\Omega _{d-2}}\int \frac{N}{2}\frac{d}{dr} \left\{
\frac{r^{d-1}}{G_{k}}\left[ F_{\gamma }(r)+\frac{1}{l^{2}}\right] ^{k}\right\} dr+B,  \label{IMini}
\end{equation}
is a functional of the fields $N:=N^{\perp }(r)f^{-2}(r)$, and $f^{2}(r)$, with $F_{\gamma}(r)
=(\gamma -f^{2}(r))/r^{2}$. The field equations obtained from (\ref{IMini}) reproduce
(\ref{EqnGravN}) and (\ref{EqnGravF}). The bulk term vanishes on the field equations, so that the
variation of the action (\ref{IMini}) on shell, is the boundary term
\begin{eqnarray}
\delta I_{T} &=&\Delta t\frac{\Sigma _{d-2}}{\Omega _{d-2}}\int \frac{d}{dr}\left(
N\frac{r^{d-1}}{2G_{k}}\delta \left[ F_{\gamma }(r)+\frac{1}{l^{2}}
\right] ^{k}\right) dr  \nonumber \\
&+&\delta B,  \label{VarIMini}
\end{eqnarray}
which means that the action is stationary on the black hole solution provided
\begin{equation}
\delta B=-\Delta tN_{\infty }\frac{\Sigma _{d-2}}{\Omega _{d-2}}\delta \mu
\;,  \label{DeltaB}
\end{equation}
and consequently, the boundary term to be added is
\[
B=-\Delta tN_{\infty }\frac{\Sigma _{d-2}}{\Omega _{d-2}}\mu +B_{0},
\]
where $B_{0}$\ is an arbitrary constant without variation. This allows identifying the mass as
\begin{equation}
M=\frac{\Sigma _{d-2}}{\Omega _{d-2}}(\mu -\mu _{0}),  \label{HMass}
\end{equation}
where the lapse at infinity ($N_{\infty }$) has been chosen equal to $1$. In order to avoid naked
singularities with positive mass, the additive constant $\mu _{0}$ is set equal to zero for all
cases except for spherical black holes solutions of Chern Simons theories ($d=2k+1$ and $\gamma
=1$), that is $\mu _{0}=$ $\frac{1}{2G_{k}}\delta _{d-2k,\gamma }.$

An alternative way to obtain the mass and angular momentum for gravity theories with
asymptotically locally AdS behavior in even dimensions ($d=2n$), has been recently proposed
\cite{Aros:1999id,Aros:1999kt}. This construction is fully covariant and background independent.
This provides an independent check of formula (\ref{HMass}), which is summarized here. The demand
on the action $I_{k}$ to have an extremum for asymptotically locally AdS spacetimes fixes the
boundary term that must be added to (\ref{Ik}) as the integral of the Euler density with a fixed
coefficient \cite{Crisostomo:2000bb},
\begin{equation}
I_{T}=I_{k}+\kappa \alpha _{n}\int {\cal E}_{2n}\;,  \label{IkEuler}
\end{equation}
where
\begin{equation}
\alpha _{n}=c_{n}^{k}:=\frac{(-1)^{n+k+1}l^{2(n-k)}}{2n\left( { {n-1 \atop k} }\right) }.
\end{equation}
The invariance of (\ref{IkEuler}) under diffeomorphisms, provides a conserved current through
Noether theorem, $d*J=0$. Assuming that the asymptotic region of the manifold is $\partial {\cal
M}=R\times \Sigma _{\gamma }$, the conserved charge associated with diffeomorphisms $x^{\mu
}\rightarrow x^{\mu }+\xi ^{\mu }$ is
\begin{equation}
Q[\xi ]=\int_{\Sigma _{\gamma }}\xi ^{\mu }\omega _{\mu }^{ab}{\cal T}_{ab},
\label{NoetherCharge}
\end{equation}
where ${\cal T}_{ab}$ is the functional derivative of the total Lagrangian in (\ref{IkEuler}) with
respect to the curvature
\begin{equation}
{\cal T}_{ab}:=\frac{\delta {L}_{T}}{\delta R^{ab}}.
\end{equation}
The mass is obtained from (\ref{NoetherCharge}) for $\xi =\partial _{t}$, without making further
assumptions about the matching with a background geometry or its topology. Thus, the mass for the
topological black holes given by (\ref{BHGeneral}) is
\begin{equation}
M=Q[\partial _{t}]=\mu \frac{\Sigma _{d-2}}{\Omega _{d-2}},
\end{equation}
in agreement with the result obtained from the Hamiltonian formalism in even dimensions.

\subsection{Entropy}

In the semiclassical approximation the partition function is given by $Z\approx e^{-I_{E}}$, where
$I_{E}$ is the Wick rotated version of the action (\ref{IMini}) given by
\begin{eqnarray}
I_{E} &=&-\beta \frac{\Sigma _{d-2}}{\Omega _{d-2}}\int_{r_{+}}^{\infty }
\frac{N}{2}\frac{d}{dr}\left\{ \frac{r^{d-1}}{G_{k}}\left[ F_{\gamma }(r)+
\frac{1}{l^{2}}\right] ^{k}\right\} dr  \nonumber \\
&+&B_{E}.
\end{eqnarray}
The on shell value of $I_{E}$ is given by $B_{E}$ and therefore the Helmholtz free energy,
$F=I_{E}/\beta =M-S/(\kappa _{B}\beta )$, is completely determined by the boundary term, where
$\left. (df^{2}/dr)\right| _{r_{+}}=4\pi \beta ^{-1}$. The boundary term $B_{E}$ is also fixed
requiring the action to have an extremum on the Euclidean form of the geometry, which covers only
the exterior section of the black hole ($r>r_{+}$). Its variation is now given by
\[
\delta B_{E}=\beta \delta M-k\frac{2\pi \Sigma _{d-2}}{\Omega _{d-2}G_{k}}
\,r_{+}^{(d-2k-1)}\left( \gamma +\frac{r_{+}^{2}}{l^{2}}\right) ^{k-1}\delta r_{+}.
\]
This implies that the variation of the entropy, as function of the horizon radius, reads
\begin{equation}
\delta S_{k}=k\frac{2\pi \kappa _{B}\Sigma _{d-2}}{\Omega _{d-2}G_{k}} \,r_{+}^{(d-2k-1)}\left(
\gamma +\frac{r_{+}^{2}}{l^{2}}\right) ^{k-1}\delta r_{+}.
\end{equation}
This relation can be integrated to yield a closed expression for entropy as a function of $r_{+}$,
given in Eq. (\ref{Entropy}).


\begin{thebibliography}{10}

\bibitem{Hawking:1983dh}
S.~W. Hawking and D.~N. Page, Commun. Math. Phys. {\bf 87}, 577 (1983).

\bibitem{Aros:1999id}
R. Aros, M. Contreras, R. Olea, R. Troncoso and J. Zanelli, Phys. Rev. Lett. {\bf 84}, 1647 (2000).

\bibitem{Aros:1999kt}
R. Aros, M. Contreras, R. Olea, R. Troncoso and J. Zanelli, Phys. Rev. {\bf D62}, 44002 (2000).

\bibitem{Abbott:1982ff}
L.~F. Abbott and S. Deser, Nucl. Phys. {\bf B195}, 76 (1982).

\bibitem{Friedman:1993ty}
J.~L. Friedman, K. Schleich, and D.~M. Witt, Phys. Rev. Lett. {\bf 71}, 1486 (1993).

\bibitem{Lemos:1995xp}
J.~P.~S. Lemos, Phys. Lett. {\bf B353}, 46 (1995).

\bibitem{Vanzo:1997gw}
L. Vanzo, Phys. Rev. {\bf D56}, 6475 (1997).

\bibitem{Brill:1997mf}
D.~R. Brill, J. Louko, and P. Peldan, Phys. Rev. {\bf D56}, 3600
(1997).

\bibitem{Aharony:1999ti}
O. Aharony, S. S. Gubser, J. Maldacena, H. Ooguri and Y. Oz, Phys. Rept. {\bf 323}, 183 (2000).

\bibitem{Witten:1998zw}
E. Witten, Adv. Theor. Math. Phys. {\bf 2}, 505 (1998).

\bibitem{Birmingham:1998nr}
D. Birmingham, Class. Quant. Grav. {\bf 16}, 1197 (1999).

\bibitem{Emparan:1999pm}
R. Emparan, C.~V. Johnson, and R.~C. Myers, Phys. Rev. {\bf D60}, 104001 (1999).

\bibitem{Balasubramanian:2000wv}
V. Balasubramanian, E. Gimon, and D. Minic, JHEP {\bf 05}, 014 (2000).

\bibitem{HDG}
R. Troncoso and J. Zanelli, Class. Quant. Grav. {\bf 17}, 4451 (2000).

\bibitem{Crisostomo:2000bb}
J. Crisostomo, R. Troncoso, and J. Zanelli, Phys. Rev. {\bf D62}, 084013 (2000).

\bibitem{LL}
C. Lanczos, Ann.Math. {\bf39}, 842 (1938); D. Lovelock, J. Math. Phys. {\bf 12}, 498 (1971).

\bibitem{Wheeler:1986nh}
J.~T. Wheeler, Nucl. Phys. {\bf B268}, 737 (1986).

\bibitem{Boulware:1985wk}
D.~G. Boulware and S. Deser, Phys. Rev. Lett. {\bf 55}, 2656 (1985).

\bibitem{Teitelboim:1987}
C. Teitelboim and J. Zanelli, Class. Quant. Grav. {\bf 4}, L127 (1987).

\bibitem{Chamseddine:1989nu}
A.~H. Chamseddine, Phys. Lett. {\bf B233}, 291 (1989); Nucl. Phys. {\bf B346}, 213 (1990).

\bibitem{Troncoso:1998va}
R. Troncoso and J. Zanelli,
Phys.\ Rev.\  {\bf D58}, 101703(R) (1998)

\bibitem{Banados:1993gq}
M. Banados, M. Henneaux, C. Teitelboim, and J. Zanelli, Phys. Rev. {\bf D48}, 1506 (1993).

\bibitem{ThanksVanzoandLouko}  We thank L. Vanzo and J. Louko, for
clarifying this point to us. See also R. Emparan, AdS/CFT duals of topological black holes and the
entropy of zero-energy states, JHEP {\bf 9906}, 036 (1999).

\bibitem{Banados:1994ur}
M. Banados, C. Teitelboim, and J. Zanelli, Phys. Rev. {\bf D49}, 975
(1994).

\bibitem{Banados:1992wn} M. Banados, C. Teitelboim, and J. Zanelli, Phys. Rev. Lett. {\bf
69}, 1849 (1992).

\bibitem{Cai:1998vy}
R. G. Cai and K. S. Soh, Phys. Rev. {\bf D59}, 044013 (1999).

\bibitem{Holzhey:1992bx}
C.~F.~E. Holzhey and F. Wilczek, Nucl. Phys. {\bf B380}, 447 (1992).

\bibitem{Susskind:1998dq}
L.~Susskind and E.~Witten, {\emph{The holographic bound in anti-de Sitter space}}, hep-th/9805114.

\bibitem{Nair:1990aa}
V.~P.~Nair and J.~Schiff, Phys.\ Lett.\ B {\bf 246}, 423 (1990).

\bibitem{Banados:1996yj}
M.~Banados, L.~J.~Garay and M.~Henneaux, Nucl.\ Phys.\ B {\bf 476}, 611 (1996)

\bibitem{Chandia:1998uf}
O. Chand\'{\i}a, R. Troncoso and J. Zanelli, {\em Dynamical Content of Chern-Simons
Supergravity},In ``Trends in Theoretical Physics II'', H. Falomir, R.E. Gamboa Saravi and F.
Schapopsnik, eds. AIP Conf. Proceedings 484,1999. Report N{${^{o}}$}: CECS-PHY-99/05, e-Print
Archive: hep-th/9903204.

\bibitem{GegenbergKunstatter}
J.~Gegenberg and G.~Kunstatter,
Phys.\ Lett.\ {\bf B478}, 327 (2000)


\bibitem{InProgress} R. Aros, C. Mart\'{\i}nez, R. Troncoso and J. Zanelli, work in progress.

\bibitem{Bautier:1997yp}
K.~Bautier, S.~Deser, M.~Henneaux and D.~Seminara,
Phys.\ Lett.\  {\bf B406}, 49 (1997)

\bibitem{Regge:1974zd}
T. Regge and C. Teitelboim, Ann. Phys. {\bf 88}, 286 (1974).



\end{thebibliography}
\end{document}